\documentclass[journal]{IEEEtran}

\usepackage{cite}
\usepackage[cmex10]{amsmath}
\usepackage[dvips]{graphicx}
\usepackage{subfigure,color}
\usepackage{float}
\usepackage{subeqnarray}
\usepackage{cases}

\usepackage{epsf}
\usepackage{amsfonts}
\usepackage{amsmath}
\usepackage{graphics,graphicx}
\usepackage{subfigure}
\usepackage{psfrag}
\usepackage{bm}
\usepackage{epsfig}
\usepackage{placeins}
\usepackage{multirow}
\usepackage{times}
\usepackage{amsmath,bm}
\usepackage{color}

\usepackage{color}

\usepackage[normalem]{ulem}

\begin{document}

\title{Polarization Control by Using Anisotropic \\3D Chiral Structures}
%
%
%

\author{Menglin~L. N.~Chen,~\IEEEmembership{Student~Member,~IEEE,}
        Li~Jun~Jiang,~\IEEEmembership{Senior~Member,~IEEE,}
        Wei~E. I.~Sha,~\IEEEmembership{Member,~IEEE,}
        Wallace~C. H.~Choy,~\IEEEmembership{Senior~Member,~IEEE,}
        and~Tatsuo~Itoh,~\IEEEmembership{Fellow,~IEEE}
\thanks{M. L. N.~Chen, L. J. Jiang, W.~E. I.~Sha and W.~C. H.~Choy are with the Department
of Electrical and Electronic Engineering, The University of Hong Kong, Hong Kong (e-mail: menglin@connect.hku.hk; jianglj@hku.hk; wsha@eee.hku.hk; chchoy@eee.hku.hk).}
\thanks{T. Itoh is with Electrical Engineering Department, University of California Los Angeles, Los Angeles, CA 90095, USA (email: itoh@ee.ucla.edu).}}


\markboth{Ieee transactions on antennas and propagation,~Vol.~XX, No.~XX, XX 2016}%
{Shell \MakeLowercase{\textit{et al.}}: Bare Demo of IEEEtran.cls for IEEE Journals}
%

\maketitle

\begin{abstract}
Due to the mirror symmetry breaking, chiral structures show fantastic electromagnetic (EM) properties involving negative refraction, giant optical activity, and asymmetric transmission. Aligned electric and magnetic dipoles excited in chiral structures contribute to extraordinary properties. However, the chiral structures that exhibit $n$-fold rotational symmetry show limited tuning capability. In this paper, we proposed a compact, light, and highly tunable anisotropic chiral structure to overcome this limitation and realize a linear-to-circular polarization conversion. The anisotropy is due to simultaneous excitations of two different pairs of aligned electric and magnetic dipoles. The 3D omega-like structure, etched on two sides of one PCB board and connected by metallic vias, achieves 60\% of linear-to-circular conversion (transmission) efficiency at the operating frequency of 9.2 GHz. The desired 90-degree phase shift between the two orthogonal linear polarization components is not only from the finite-thickness dielectric substrate but also from the anisotropic chiral response slightly off the resonance. The work enables elegant and practical polarization control of EM waves.
\end{abstract}


\section{Introduction}

\IEEEPARstart{C}{hiral} structures are composed of particles that cannot be superimposed on their mirror images. The asymmetric geometry feature of a chiral particle results in the cross coupling between electric field and magnetic field. Therefore, a chiral medium is also known as a bi-isotropic medium if it has identical electromagnetic responses in all directions~\cite{kong}. Chiral media can be found in nature. However, the chirality is usually very weak. The artificial materials can enhance the needed chiral properties. With the strong chirality, the chiral structure could easily realize negative refractive index compared to the conventional negative-index metamaterial composed of split-ring resonator (SRR) and metallic wires~\cite{pendry2004chiral}. Besides the negative refractive index, chiral structure shows other interesting features like giant optical activity, circular dichroism and asymmetric transmission~\cite{wang2009chiral,li2013chiral,menzel2010asymmetric}.

Various man-made chiral "molecules" have been analyzed and the corresponding parameter retrieval method has been studied~\cite{retrieval}. Generally, chiral structure can be classified into two groups: planar chiral structure and three-dimensional (3D) chiral structure. Planar chiral structures like rosettes~\cite{rosettes,rogacheva2006giant} and cross wires~\cite{cross} are easy to fabricate. They exhibit giant optical activity and negative refractive index at different frequency bands for right circular polarized (RCP) and left circular polarized (LCP) waves. Typical prototypes of the 3D chiral particle originate helical geometry, such as chiral SRR~\cite{nonplanarz}, omega-shaped particle~\cite{omega,zhang}, and cranks~\cite{crank}. These 3D chiral particles have been well designed so that they can be fabricated as planar structures on PCBs. Moreover, the super-cell technique has been applied in the U-shaped chiral particles~\cite{u,u-cir} to gain more flexible tuning. Among all the chiral particles, planar chiral particles usually present $n$-fold rotational symmetry. The rotational symmetry makes the planar chiral structure insensitive to the polarization direction of normal incidence waves and sets up limitations to its polarization tuning properties. Unlike the planar chiral structure, the 3D chiral structure is sensitive to the polarized state of incident wave and allows for a flexible control of polarization. Therefore, the polarization conversion such as linear-to-circular conversion can be realized by the 3D chiral structure.

In this paper, we explored the polarization properties of a 3D omega-shaped chiral structure comprehensively. Even though a similar structure has been proposed in~\cite{zhang} and its origin of chirality is explained by a resonant LC circuit. We explore the physical origin of chirality by the 3D omega-shaped structure based on the induced electromagnetic fields. And then, we offer a new physical insight to the excitation condition and to the polarization responses with varying geometries. Here, we found the 3D omega-shaped chiral structure shows a great capability to manipulate the polarization state of electromagnetic waves. First, we show the transmitted polarization state by the chiral structure can be tuned in a wide range by twisting the arms of the chiral particle. Second, we theoretically and experimentally demonstrate an anisotropic omega-shaped chiral structure that functions as a circular polarizer. Compared to conventional polarizers, the chiral polarizer has an ultra-compact volume. Third, we designed a uniaxial omega-shaped chiral structure. It generates giant optical activity which is not sensitive to the polarization state for normal incidence waves and shows advantages over other 3D chiral structures arranged in the Bravais lattice~\cite{nonplanarz}. All the chiral structures can be conveniently fabricated using the PCB technique, measured, and compared with simulations.

\section{Origin of chirality}
In this section, we discuss the origin of the chirality of the 3D omega-shape by analyzing the directions of the induced electromagnetic fields, that was not published before, according to out best knowledge.

\subsection{ME dipole pairs}
\begin{figure}[!tbc]
\centering
\subfigure[]{
\includegraphics[width=0.4\columnwidth]{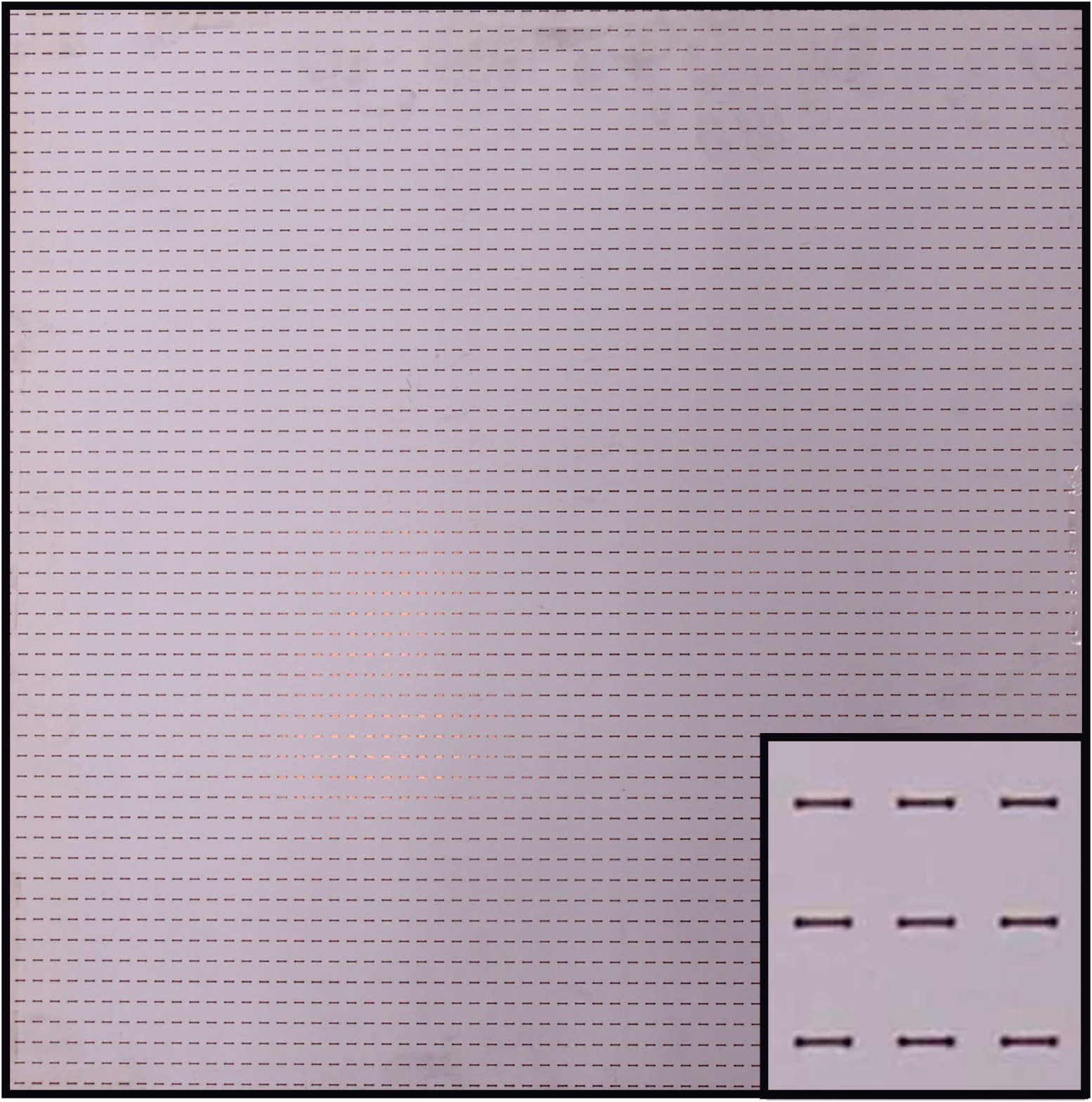}}
\subfigure[]{
\includegraphics[width=0.4\columnwidth]{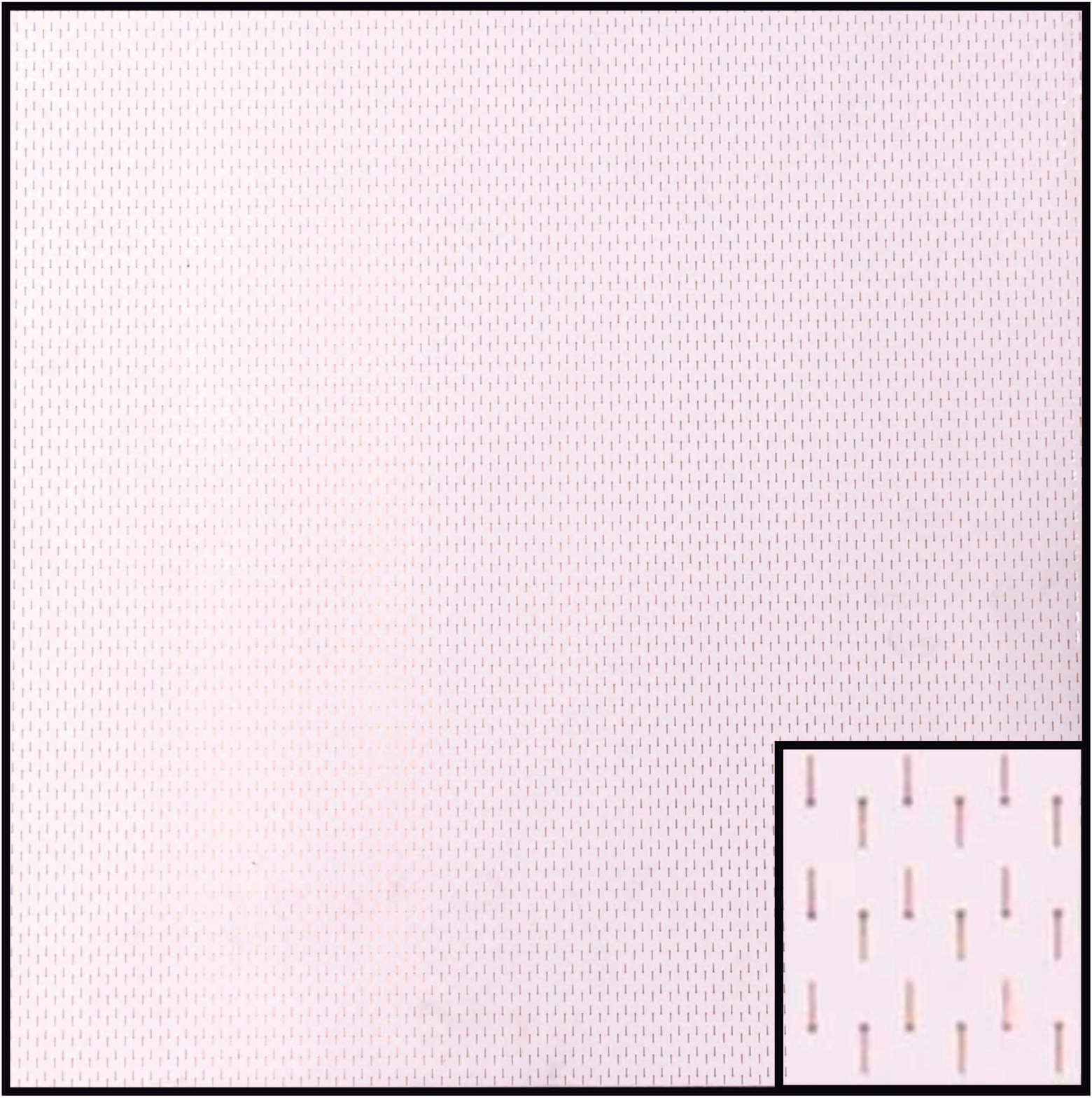}}
\subfigure[]{
\includegraphics[width=0.4\columnwidth]{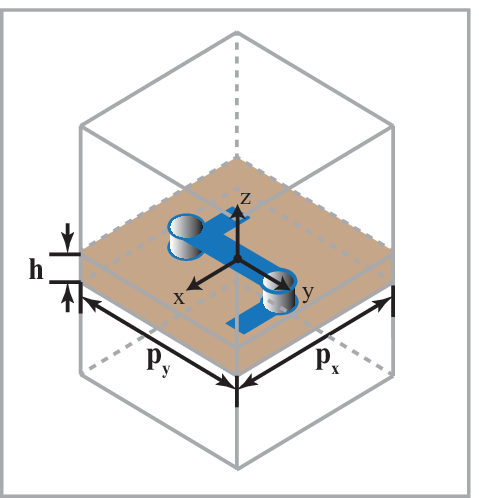}}
\subfigure[]{
\includegraphics[width=0.4\columnwidth]{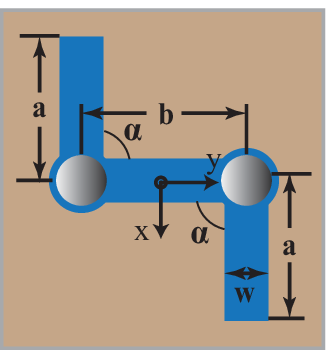}}
\caption{Illustration of the chiral structure. (a) photograph of the top layer of a fabricated sample slab; (b) photograph of the bottom layer of the fabricated sample slab; (c) schematic of the twisted omega-like chiral unit cell with the periodicity along the $x$ and $y$ direction; (d) top view of the chiral unit cell. Lattice constants are denoted by $p_x$ and $p_y$, respectively. The thickness of dielectric substrate is $h$, which is the same as the height of the two vias of the chiral structure. Arm lengths of the unit cell is $a$ and $b$ at the bottom and the top layers, respectively. The angle between the arm at the top layer and that at the bottom layer is represented by $\alpha$.}
\label{photo}
\end{figure}

The proposed chiral structure is shown in Fig.~\ref{photo}. Figure~\ref{photo}(c) and (d) show the schematic pattern of one chiral particle at different viewing angles. The chiral particle can be simply seen as a twisted conducting wire in 3D domain. It consists of five segments: the two vertical segments (vias) connect the two horizontal segments (wires) and one vertical segment (wire), which are placed at the bottom and top of the substrate, respectively. This geometry has a complete symmetry breaking along the $x$, $y$ and $z$ directions, indicating the strong chirality.

\begin{figure}[!tbc]
\centering
\subfigure[]{
\includegraphics[width=0.3\columnwidth]{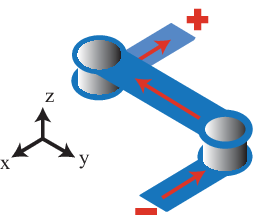}}
\subfigure[]{
\includegraphics[width=0.3\columnwidth]{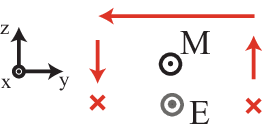}}
\subfigure[]{
\includegraphics[width=0.3\columnwidth]{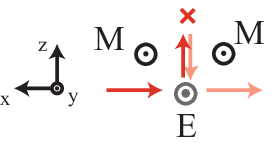}}
\caption{Illustration of fundamental mode of the proposed omega-like chiral unit cell. The surface current (red arrow) and generated magnetic and electric dipoles are observed at different viewing angles. (a) 3D view of the chiral particle resonating at the (half-wavelength) fundamental mode; (b) current distribution viewed from the $x$ axis and the induced ME dipole pair along the $x$ direction; (c) current distribution viewing from the $y$ axis and the induced ME dipole pair along the $y$ direction.}
\label{current}
\end{figure}

The connected chiral particle has the total length of $l$, where $l=2a+b+2h$. $a$ and $b$ are the lengths of the segments shown in Fig.~\ref{photo}(d) and $h$ is the height of the vertical segments. The first (fundamental) resonance of the twisted wire is the half-wavelength resonance, which should satisfy $l=\lambda_{eff}/2$, where $\lambda_{eff}$ is the effective wavelength. In the following analysis, we only consider the half-wavelength fundamental mode to achieve a compact design. Under this condition, the periodic length (lattice constant) of the chiral structure is much smaller than the incident wavelength, so that only the zeroth-order diffraction exists. Current direction and charge distribution are drawn in Fig.~\ref{current}(a). Charges are accumulated at the two ends of the wire, forming an electric dipole (E dipole) in the $xoy$ plane. The E dipole pointed toward the two ends has both the $x$ and $y$ components. By looking at the $x$ direction as shown in Fig.~\ref{current}(b), a current loop can be formed and it generates a magnetic dipole (M dipole) pointing at the $x$ direction. The M dipole is aligned with the $x$ component of the E dipole. Besides the ME dipole pair, when we look at the $y$ direction, another ME dipole pair aligned with the $y$ direction can be found as illustrated in Fig.~\ref{current}(c). It is well known that chirality of a medium arises from the coupled electric and magnetic fields. In the proposed chiral particle, there are two anisotropic pairs of coupled ME dipoles. This special configuration facilitates the tunable chirality that the strengths of the two components of the E (M) dipoles can be adjusted by changing the angle $\alpha$ and segment lengths (See Fig.~\ref{photo}(d)).

\subsection{Excitation}
\begin{figure}[!tbc]
\centering
\subfigure[]{
\includegraphics[width=0.4\columnwidth]{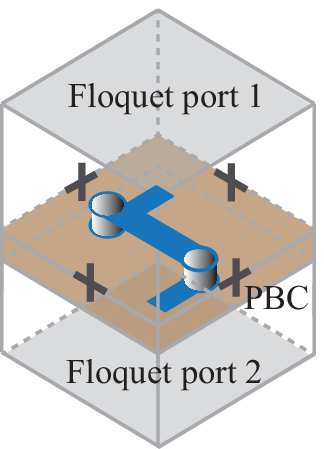}}\hspace{1em}
\subfigure[]{
\includegraphics[width=0.4\columnwidth]{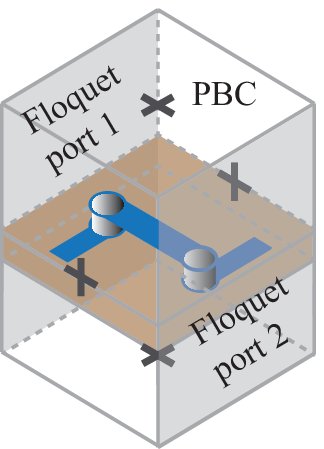}}
\caption {Simulation configurations in HFSS. (a) excitation from top with two pairs of periodic boundaries at the four lateral faces; (b) excitation from the side with one pair of periodic boundaries on the two lateral faces and the other pair on the top and bottom faces. Here, $++++$ stands for the periodic boundary.}
\label{setup}
\end{figure}

Excitation condition for the fundamental resonant mode of the omega-like chiral particle is explored. One unit cell in a cuboid box with two sets of periodic boundaries and one set of Floquet port is simulated in Ansoft HFSS as shown in Fig.~\ref{setup}.

When the periodicity is along the $x$ and $y$ axes, i.e. periodic boundary conditions (PBC) are applied at the transverse $xoy$ plane while two floquet ports are assigned on the top and bottom boundaries along the $z$ direction, the fundamental mode can be successfully excited. As discussed, we have the $x$ and $y$ components for both M dipole and E dipole. When the plane wave propagates along $z$ axis, no matter what the polarization direction is, both incident electric field and incident magnetic field could be aligned with the corresponding E dipole and M dipole. As a result, the mode conversion occurs between the plane wave (propagating in free space) and the fundamental standing wave (supported in the chiral particle)~\cite{chew}. Due to the bi-anisotropy in this chiral particle, the spatial overlap between the plane wave and the fundamental mode highly depends on the polarization state of the plane wave. Therefore, although the chiral particle can be excited under either $x$ or $y$ polarized incident field, the two excitations have different transmission and reflection responses.

If the periodicity is along the $z$ and $x$ axes, the fundamental mode cannot be excited due to the polarization misalignment. When the wave impinging from the lateral side is $x$ polarized, the z-polarized H field is not aligned with the M dipole, since there is no z component of the M dipole. When the wave is polarized at the $z$ direction, the E dipole at the $xoy$ plane cannot be aligned with the incident E field. Similarly, the fundamental mode cannot be excited when the periodicity is along the $z$ and $y$ axes. In conclusion, the whole chiral structure can only be significantly excited by the normal incidence wave along the $z$ axis.

From above analyses, the chiral slab could be realized by using the PCB technique with a substrate inserted into the $xoy$ plane. Moreover, it is important to emphasize that the induced ME dipole pairs are parallel to the substrate plane, which is quiet different from the 3D chiral particle reported in~\cite{nonplanarz}. The photograph of a fabricated chiral slab is shown in Fig.~\ref{photo}(a) and (b).

\section{Polarization control}

Assuming a plane wave propagates along the $z$ direction and penetrates a chiral medium, the incident and transmitted $E$ field can be decomposed into the two linear $x$ and $y$ components,

\begin{equation}
\mathbf{E}_i(\mathbf{r},t)=\begin{pmatrix}
                            i_x \\ i_y
                           \end{pmatrix}
                           e^{i(kz-\omega t)},\quad
                           \mathbf{E}_t(\mathbf{r},t)=\begin{pmatrix}
                            t_x \\ t_y
                           \end{pmatrix}
                           e^{i(kz-\omega t)}
\end{equation}
where $\omega$ is the wave frequency, $k$ is the wave number, and the complex amplitudes $i_x$, $i_y$ and $t_x$, $t_y$ are polarization states for incident and transmitted waves.

To model a chiral particle, the transmission matrix $T$, that connects the polarization state of the transmitted wave to that of the incident wave is constructed in the linear basis~\cite{PRA_Jones_Calculus}:

\begin{equation}
\begin{pmatrix}
  t_x \\ t_y
\end{pmatrix}
=
\begin{pmatrix}
    T_{xx} & T_{xy} \\
    T_{yx} & T_{yy}
  \end{pmatrix}
\begin{pmatrix}
  i_x \\ i_y
\end{pmatrix}
=T_{lin}
\begin{pmatrix}
  i_x \\ i_y
\end{pmatrix}
\end{equation}
where the first and second subscripts of $T$ denote the polarization states of the transmitted and incident waves, respectively. Then, the transmission matrix in the circular basis can be obtained from that in the linear basis (Eq.~3).

\begin{figure*}[!t]
\normalsize
\setcounter{equation}{2}
\begin{equation}
T_{circ}=
\begin{pmatrix}
    T_{++} & T_{+-} \\
    T_{-+} & T_{--}
  \end{pmatrix}
=
  \frac{1}{2}
  \begin{pmatrix}
    (T_{xx}+T_{yy})+i(T_{xy}-T_{yx}) & (T_{xx}-T_{yy})-i(T_{xy}+T_{yx})\\
    (T_{xx}-T_{yy})+i(T_{xy}+T_{yx}) & (T_{xx}+T_{yy})-i(T_{xy}-T_{yx})
  \end{pmatrix}
\label{circ}
\end{equation}
where $+$ and $-$ represent the RCP and LCP waves.
\hrulefill
\vspace*{4pt}
\end{figure*}

For a bi-isotropic chiral medium, the coupled electric and magnetic fields result in two different eigensolutions for plane waves with two eigenvectors corresponding to the RCP wave and LCP wave, respectively. Thus, the polarization of incident wave changes through the chiral medium. Polarization properties of a chiral structure are characterized by the optical activity and circular dichroism. Optical activity stands for the polarization rotation phenomenon for a linearly polarized incident wave. Mathematically, it is represented by the azimuthal rotation angle $\theta$. Circular dichroism characterizes the polarization transition of waves. For example, linear polarization changes to elliptical one. The circular dichroism is measured by the ellipticity $\eta$. $\theta$ and $\eta$ can be calculated by

\begin{subequations}
\begin{align}
\theta &=\frac{1}{2}[\text{arg}(T_{++})-\text{arg}(T_{--})],
\\
\eta &=\frac{1}{2}\sin^{-1}\left(\frac{|T_{++}|^2-|T_{--}|^2}{|T_{++}|^2+|T_{--}|^2}
\right)
\end{align}
\end{subequations}

For planar chiral structures, typically having three-fold ($C_3$) or four-fold ($C_4$) rotational symmetry, the transmission matrix has the following forms,

\begin{subequations}
\begin{align}
T_{lin}^{C_4} &=
\begin{pmatrix}
    T_{xx} & T_{xy} \\
    -T_{xy} & T_{xx}
  \end{pmatrix},
\\
T_{circ}^{C_4} &=
  \begin{pmatrix}
    T_{xx}+iT_{xy} & 0\\
    0 & T_{xx}-iT_{xy}
  \end{pmatrix}
\end{align}
\end{subequations}

It can be seen that the transmission coefficients in linear basis are not independent but show specific relations. The resultant transmission matrix in circular basis is diagonal.

\begin{figure}[!tbc]
\centering
\subfigure[]{
\includegraphics[width=0.4\columnwidth]{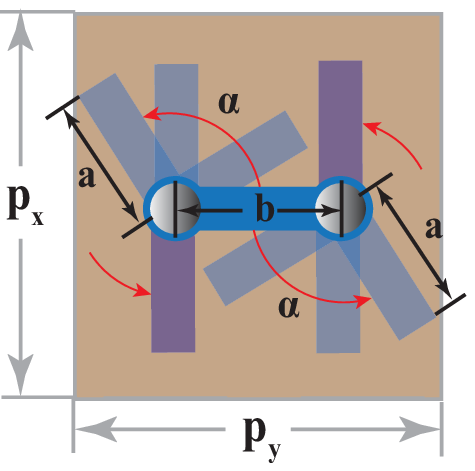}}
\subfigure[]{
\includegraphics[width=0.2\columnwidth]{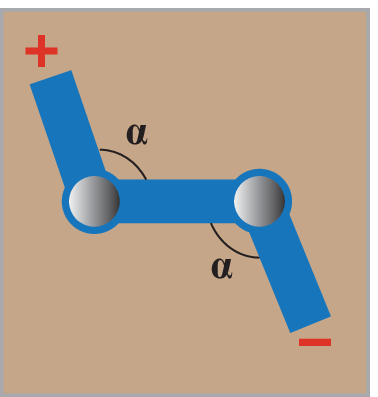}}
\subfigure[]{
\includegraphics[width=0.2\columnwidth]{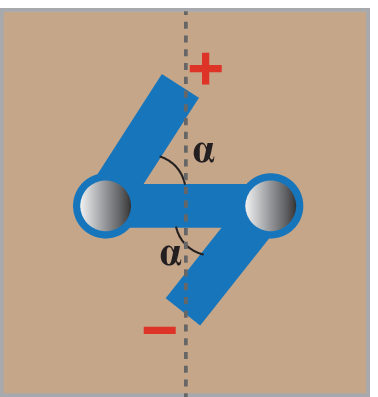}}
\subfigure[]{
\includegraphics[width=0.2\columnwidth]{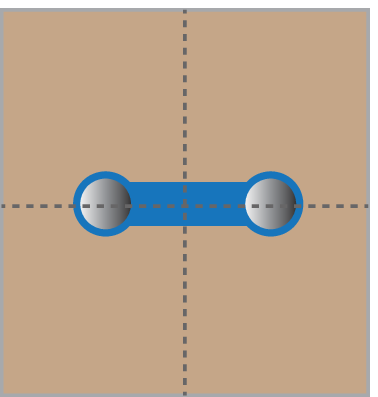}}
\subfigure[]{
\includegraphics[width=0.2\columnwidth]{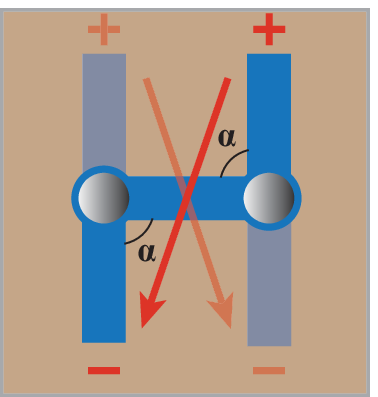}}
\caption{Schematic of the chiral unit cell with different configurations of twisting angle $\alpha$. The signs of accumulated charges at the two ends of the chiral unit cell are denoted under the illumination of the $x$ polarized wave. (a) the rotation pattern of the two arms at the horizontal plane; (b) $\alpha>90^{\circ}$; (c) $\alpha<90^{\circ}$; (d) $\alpha=0^{\circ}$; (e) $\alpha=-90^{\circ}$ (solid blue color) and $\alpha=90^{\circ}$ (transparent blue color). Arm lengths are set to be $a=b=3$~mm. The lengths of the vertical vias is $h=1.6$~mm. The square unit cell occupies $8\times8$~mm.}
\label{alpha}
\end{figure}

Through our design, optical activity and circular dichroism can be engineered by tailoring the mutual coupling between the E and M dipoles. Since there are many degrees of freedom for the proposed chiral unit cell including the segment lengths $a$ and $b$, the height of the vias $h$, and the twisting angle $\alpha$ [see Fig.~\ref{photo}(d)], the azimuthal rotation angle and ellipticity could be tunable over a large range. Here, we tune the chiral property by modifying $\alpha$ from positive values to zero then to negative values as illustrated in Fig.~\ref{alpha}.

The angle $\alpha$ greatly influences the direction and strength of the induced E and M dipoles. For example, when $\alpha$ increases in the first two quadrants, as depicted in Fig.~\ref{alpha}(a), the separation between the two ends of the chiral unit cell increases. Consequently, the strength of the E dipole decreases, and the coupling between the E dipole and M dipole is weaken. Chirality depending on the coupling between ME dipoles will be reduced in Fig.~\ref{alpha}(b) comparing to Fig.~\ref{alpha}(c). Moreover, $\alpha$ also determines the direction of ME dipoles. For example, in Fig.~\ref{alpha}(c), no $y$ component of the induced E dipole can be found in the chiral unit cell. In Fig.~\ref{alpha}(d), when $\alpha=0$, the direction of the induced E dipole only has the $y$ component; and the M dipole only has the $x$ component. In this case, no aligned ME dipole pair can be generated, resulting in the vanishing chirality. Additionally, when $\alpha$ goes to negative values, for instance $-90^{\circ}$ in Fig.~\ref{alpha}(e), compared to the case of $\alpha=90^{\circ}$, the strengthes of the induced fields are identical, but the direction of the $y$ component of the induced E dipole is reversed. Thus, the cross transmission coefficients of $T_{xy}$ and $T_{yx}$ have opposite signs for the cases of $\alpha=90^{\circ}$ and $\alpha=-90^{\circ}$. Mathematically, we have,

\begin{equation}
T_{lin}^{\alpha}=
\begin{pmatrix}
    T_{xx} & T_{xy} \\
    T_{yx} & T_{yy}
  \end{pmatrix},
T_{lin}^{-\alpha}=
\begin{pmatrix}
    T_{xx} & -T_{xy} \\
    -T_{yx} & T_{yy}
  \end{pmatrix}
\end{equation}

\begin{equation}
T_{circ}^{\alpha}=
\begin{pmatrix}
    T_{++} & T_{+-} \\
    T_{-+} & T_{--}
  \end{pmatrix},
T_{circ}^{-\alpha}=
\begin{pmatrix}
    T_{--} & T_{-+} \\
    T_{+-} & T_{++}
  \end{pmatrix}
\end{equation}
We can derive that $\theta^{\alpha}=-\theta^{-\alpha}$ and $\eta^{\alpha}=-\eta^{-\alpha}$.

\begin{figure}[!tbc]
\centering
\psfrag{m}[c][c][0.6]{$\bm{\theta}$ \textbf{(deg)}}
\psfrag{n}[c][c][0.6]{$\bm{\eta}$ \textbf{(deg)}}
\psfrag{c}[l][c][0.5]{$\bm{\alpha}= 0^\circ$}
\psfrag{h}[l][c][0.5]{$\bm{\alpha}= 30^\circ$}
\psfrag{d}[l][c][0.5]{$\bm{\alpha}= 60^\circ$}
\psfrag{f}[l][c][0.5]{$\bm{\alpha}= 90^\circ$}
\psfrag{e}[l][c][0.5]{$\bm{\alpha}=-90^\circ$}
\psfrag{g}[l][c][0.5]{$\bm{\alpha}= 130^\circ$}
\subfigure[]{
\includegraphics[width=0.48\columnwidth]{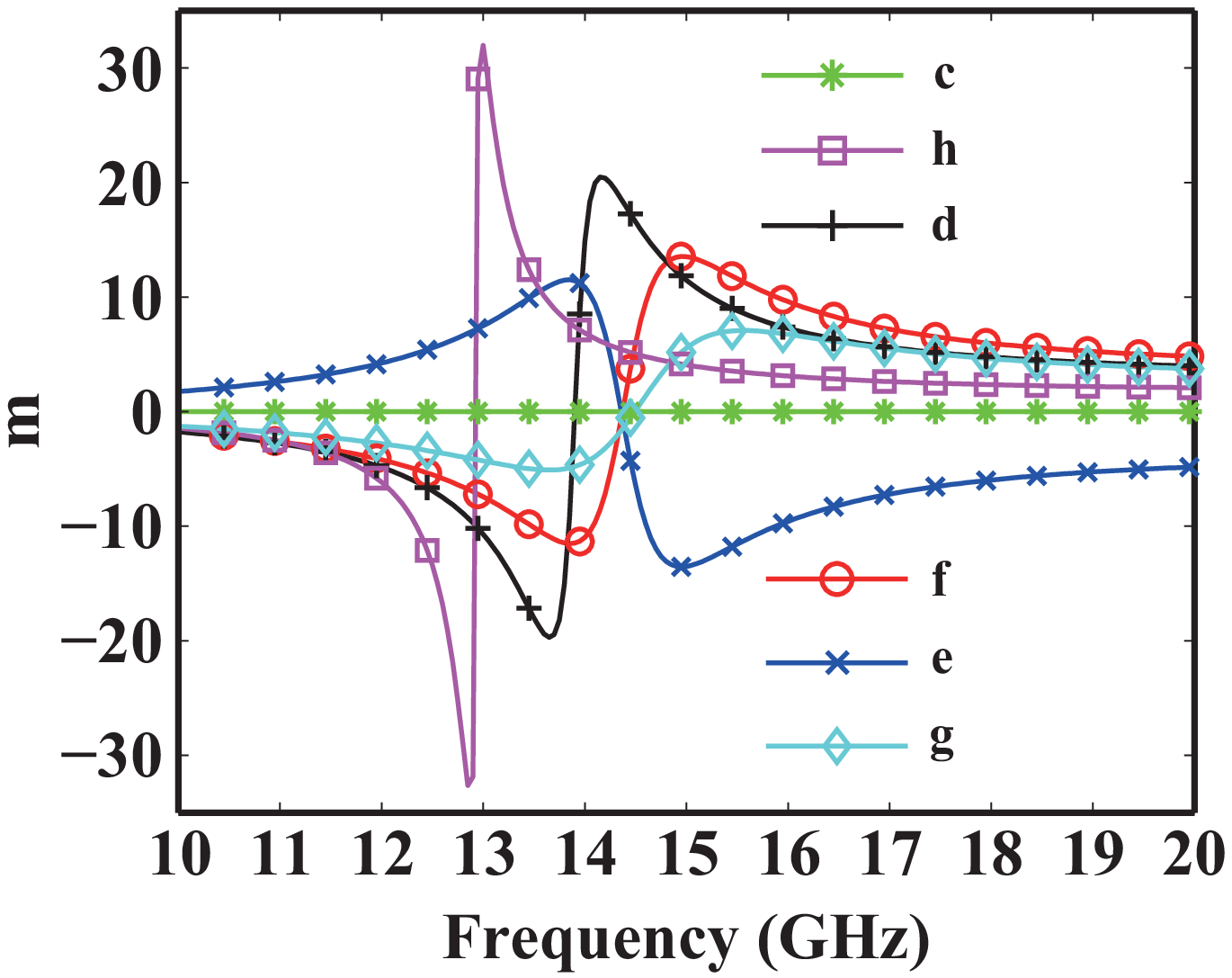}}
\subfigure[]{
\includegraphics[width=0.48\columnwidth]{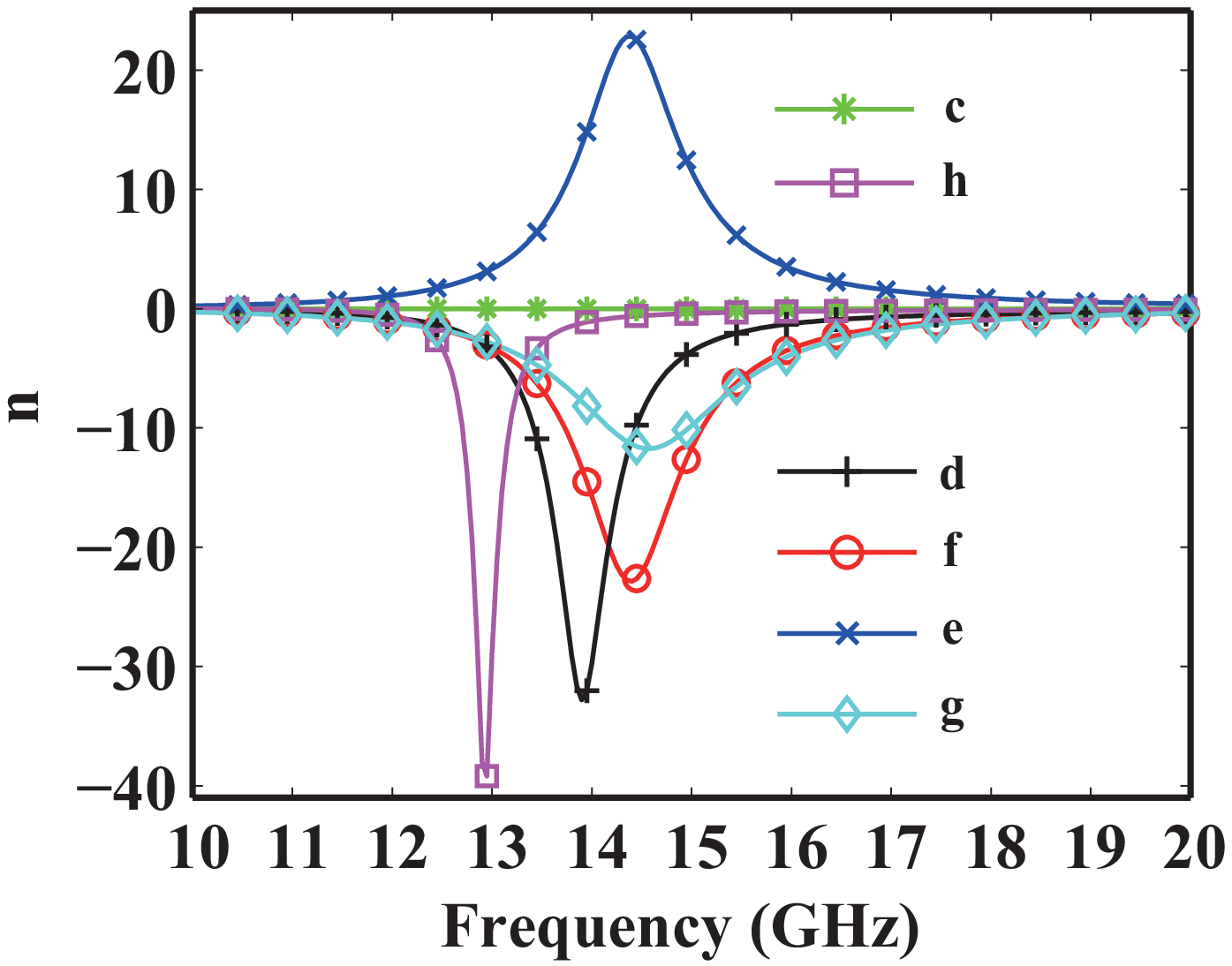}}
\caption{Tunable polarization properties of the chiral structure with a modified twisting angle $\alpha$. The other parameters are the same as described in Fig.~\ref{alpha} and no dielectric substrate is adopted. (a) azimuthal rotation angle; (b) ellipticity.}
\label{result_alpha}
\end{figure}

For simplicity, no dielectric substrate is considered in the simulation, which does not affect the conclusion to be made. The azimuthal rotation angle and ellipticity as a function of the twisting angle $\alpha$ are plotted in Fig.~\ref{result_alpha}. As expected, the azimuthal rotation angle and ellipticity are zeros in the whole frequency band when $\alpha=0$ (the green-star curves). When $\alpha$ is not equal to zero, the peak values of both $\theta$ and $\eta$ increase as $\alpha$ decreases. The azimuthal angles and ellipticity have opposite signs for $\alpha= 90^{\circ}$ and $\alpha= -90^{\circ}$ cases. All the simulation results are in good agreement with above theoretical analyses. It is worthy of noticing that the resonant frequency of the chiral particle is shifted to a lower frequency as $\alpha$ decreases. This can be explained by the influence of $\alpha$ on the mutual coupling between the ME dipoles. Stronger coupling of the ME dipoles can be regarded as extra LC loads of the chiral particle leading to a lower resonant frequency. When $\alpha$ is larger than $90^{\circ}$, the coupling of the ME dipoles has already become very weak so that its influence becomes less obvious. Meanwhile, as $\alpha$ keeps increasing, the adjacent ends of the two cells carrying opposite electric charges are getting closer. This provides extra LC loads leading to the lower resonant frequency. On the other hand, increasing $\alpha$ makes the two ends of a single cell further. These two effects both play a role in determining the resonant frequency, so there is no apparent difference between the two resonant frequencies when $\alpha= 90^{\circ}$ and $\alpha= 130^{\circ}$.

\section{Circular polarizer implementation}

To convert a linearly polarized wave to a circularly polarized wave, a birefringent material is needed, such as the metasurface in~\cite{cui}. Lack of any symmetry, our proposed 3D chiral structures with large circular dichroism can be engineered to realize the linear to circular polarization conversion.

\subsection{Simulations}
The two pairs of aligned ME dipoles cause the anisotropy of the proposed chiral structure. Due to the anisotropy and highly tunable feature, the chiral structure can be designed for converting an $x$ polarized wave to a circularly polarized wave. In this design, the dielectric substrate is chosen as AD600. The permittivity and thickness of the dielectric substrate is $\epsilon_r=6.15$ and $h=1.524$~mm. The loss tangent of the material is $0.003$. Segment lengths $a$ and $b$ are carefully optimized. The final geometrical parameters for the unit cell in Fig.~\ref{photo}(c) and (d) are: the lengths of wire segments $a=2.9$~mm, $b=2.5$~mm, the radii of vias are $0.2$~mm, the width of line segments is $0.4$~mm. $\alpha=90^\circ$ and the period of the unit cell is $p_x=7$~mm and $p_y=6$~mm. For experiments, we fabricated a chiral sample with $54 \times 63$ unit cells. The sample occupies an overall area of $378 \times 378$~mm$^{2}$ as shown in Fig.~\ref{photo}(a) and (b).

\begin{figure}[!tbc]
\centering
\psfrag{a}[l][c][0.6]{$\bm{\alpha=90^{\circ}}$}
\psfrag{c}[l][c][0.5]{$\bm{T_{xx}}$ (Sim.)}
\psfrag{e}[l][c][0.5]{$\bm{T_{yx}}$ (Sim.)}
\psfrag{g}[l][c][0.5]{$\bm{T_{xx}}$ (Meas.)}
\psfrag{h}[l][c][0.5]{$\bm{T_{yx}}$ (Meas.)}
\psfrag{d}[l][c][0.5]{$\bm{90^{\circ}}$}
\psfrag{b}[l][c][0.6]{$\bm{\alpha=-90^{\circ}}$}
\psfrag{f}[l][c][0.5]{$\bm{T_{yx}}$ (Sim.)}
\psfrag{i}[l][c][0.5]{$\bm{F_{xx}}$ (Sim.)}
\psfrag{j}[l][c][0.5]{$\bm{F_{yx}}$ (Sim.)}
\subfigure[]{
\includegraphics[width=0.48\columnwidth]{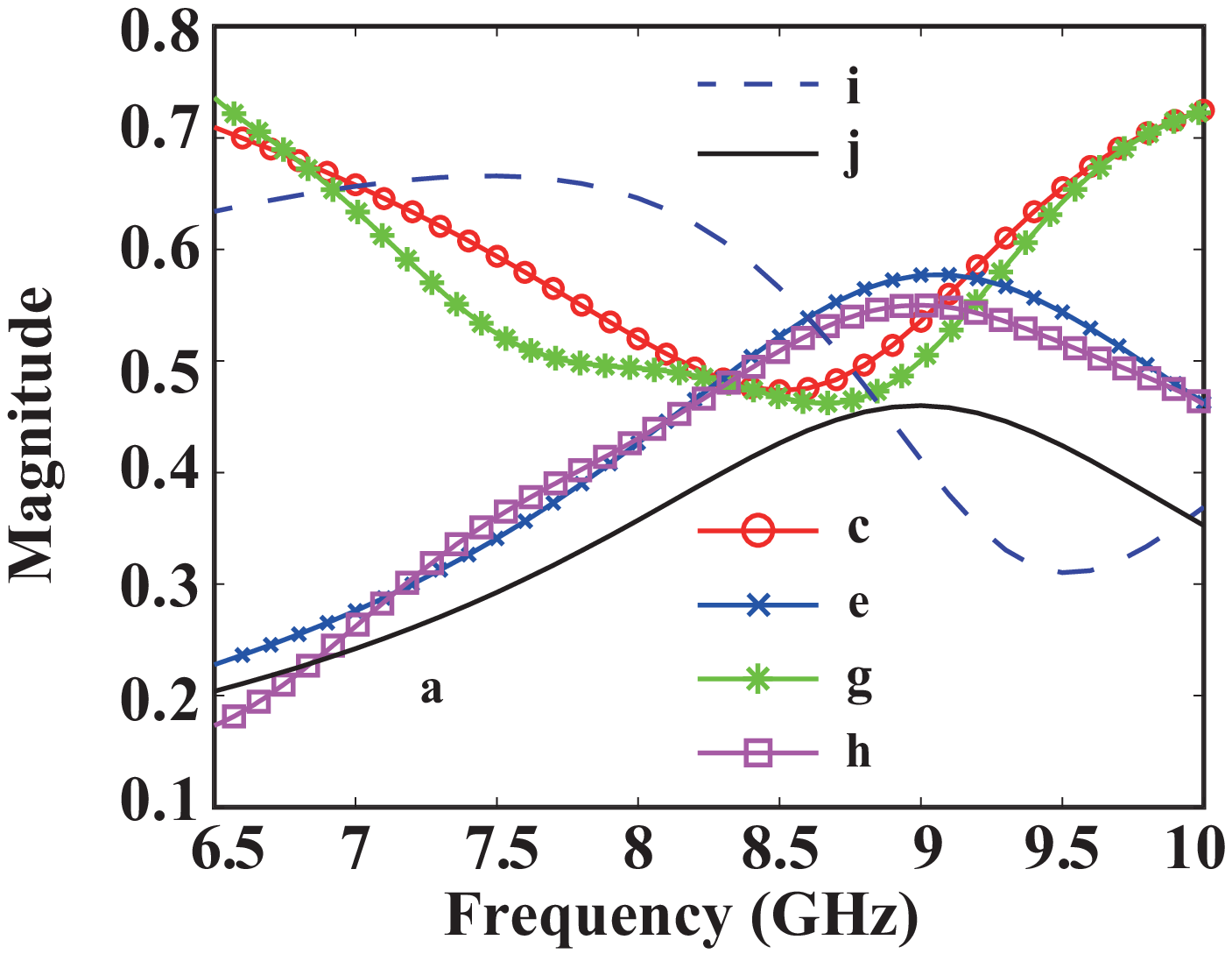}}
\subfigure[]{
\includegraphics[width=0.48\columnwidth]{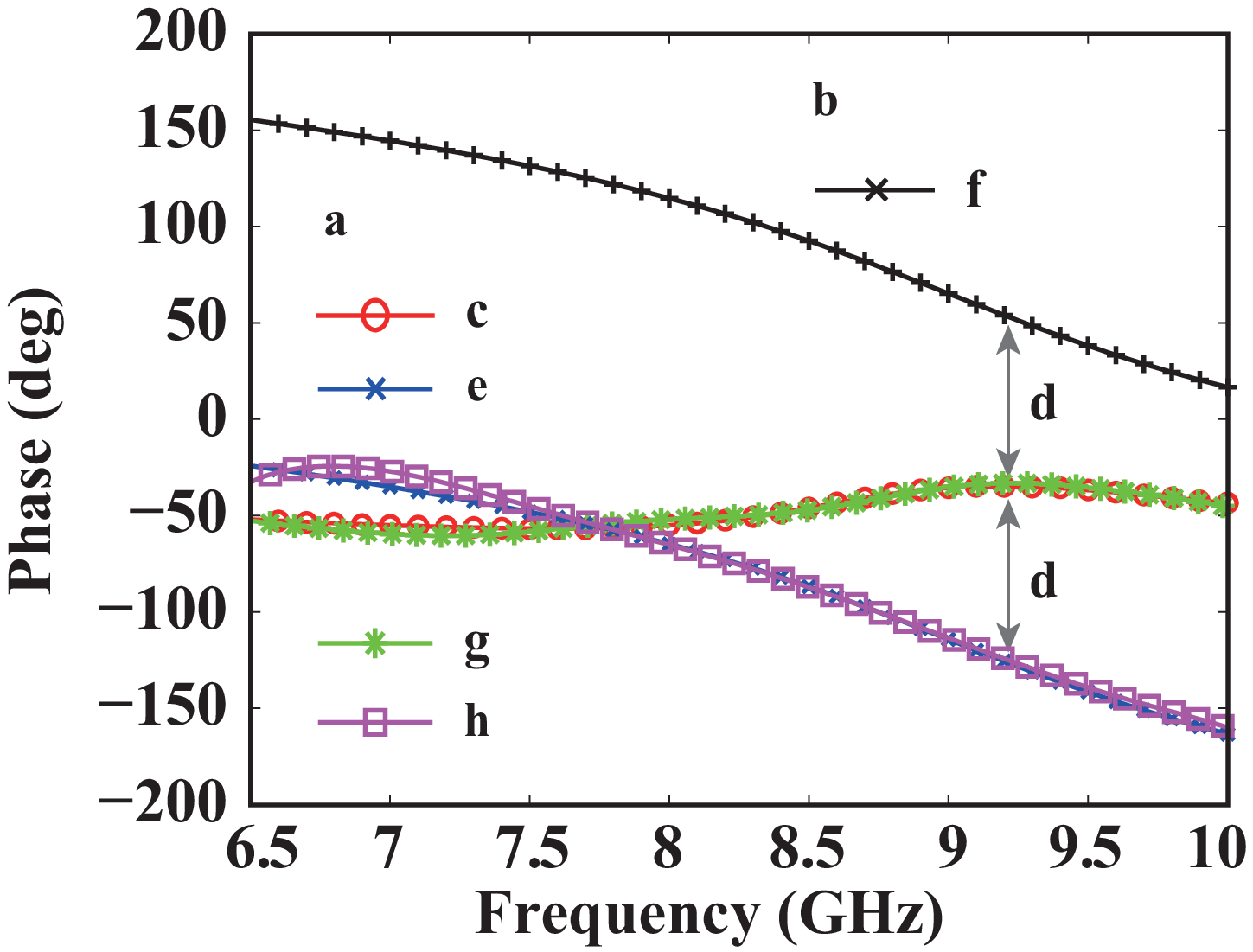}}
\caption{Simulation and experimental results of linear transmission coefficients for the chiral circular polarizer when illuminated by a $x$ polarized wave. (a) magnitude of transmission coefficients; (b) phase of transmission coefficients.}
\label{meas_circ}
\end{figure}

The phase delay between the transmitted $x$ and $y$ components is adjusted based on two factors: 1) arm lengths $a$ and $b$, as depicted in Fig.~\ref{photo}(d). Similar to the twisting angle $\alpha$, $a$ and $b$ also influence the direction and strength of the induced E and M dipoles. 2) chiral anisotropy at off-resonant frequencies to guarantee the same amplitude and desired retardation for cross polarized components. Simulation and experiment results are shown in Fig.~\ref{meas_circ}(a) and (b). Within the whole measured frequency range, a reasonable agreement between simulation and measurement results can be observed. At the operating frequency of $9.2$~GHz, the magnitudes of the transmitted $E_x$ and $E_y$ components are equal to $0.565$. With respect to the transmitted $Ey$ component, the phase of the transmitted $E_x$ component is retarded by $90^\circ$, indicating a transmitted RCP wave. The efficiency of a circular polarizer is determined by many factors, such as the substrate loss, copper loss, matching property and undesired cross-polarized component. Compared to existing chiral polarizers, our proposed one has a great advantage that the phase delay between the transmitted $x$ and $y$ components is $90^\circ$, which attributes to the large chirality and high tunability. Therefore, there is no unwanted cross-polarized wave, i.e. the LCP one in this case. A portion of the incident wave is reflected back due to the mismatch between the chiral slab and air interface. As can be found in Fig.~\ref{meas_circ}(a), at the operating frequency, the total reflected wave occupies $33\%$ of the total incident energy. With the loss of the material counted, the remaining is completely converted to the RCP wave. From the experimental results, the conversion efficiency of the chiral polarizer is about $64\%$. In contrast to conventional polarizers, the chiral polarizer has an ultra-compact design. The size of the chiral unit cell at the operating frequency is approximated to be $0.21\lambda_0 \times 0.18\lambda_0$, where $\lambda_0$ is the incident wavelength.

Furthermore, based on previous descriptions, we can obtain a $180^{\circ}$ phase shift for $T_{yx}$ by simply switching the two arm orientations (twisting angle $\alpha$ is changed from $90^{\circ}$ to $-90^{\circ}$). Interestingly, the switched chiral polarizer can convert the $x$ polarized wave to a LCP wave instead of RCP wave, as presented in Fig.~\ref{meas_circ}(b).

\subsection{Experiments}

\begin{figure}[!tbc]
\centering
\includegraphics[width=0.8\columnwidth]{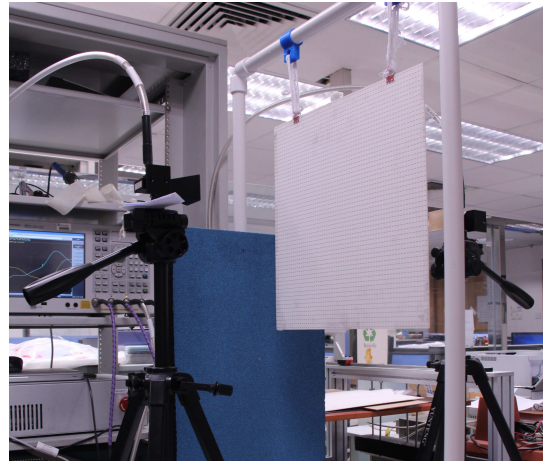}
\caption{Experimental setup for the transmission measurements of periodic chiral structures.}
\label{experiment}
\end{figure}

Measurements are implemented via a free-space electromagnetic transmission system. Two standard linear-polarized horn antennas working at the frequency ranging from $6.57$~GHz to $9.99$~GHz are set as a transmitter and receiver, respectively, as shown in Fig.~\ref{experiment}. A vector network analyzer (VNA) is used to record and process time-domain transmitted signals. Since our horn antennas only emit and receive linearly polarized waves, transmission coefficients in linear basis are obtained first. Circular transmission coefficients are then calculated based on the linear ones by Eq.~\ref{circ}.

For co-transmission coefficients, two horn antennas need to be aligned and the electromagnetic response between them are calibrated. In our case, the distance between the two horn antennas is chosen to be around $60$~cm to 1) make sure the wave impinging on the sample is a plane wave; 2) avoid the edge/truction effect of the finite periodic structures; 3) guarantee that sufficient unit cells are illuminated. Next, the sample is inserted between the two antennas. Cross-transmission coefficients are measured by rotating the receiving horn antenna by $90$ degrees.

During the experiment, a time-domain gate technique is employed to eliminate the disturbances from the mismatch of antennas and multiple reflections between the antennas and sample. Gate parameters are first estimated with the distance from the sample to the receiver and transmitter. Then, the gate parameters are carefully tuned and chosen. After incorporating the time-domain gate, unwanted echoes are eliminated resulting in a smoother response in the frequency domain.

Measurement results are shown in Fig.~\ref{meas_circ}. They are in good agreements with the simulation ones. The phase difference between $T_{xx}$ and $T_{yx}$ at the operating frequency is measured to be $90^{\circ}$; and magnitudes of both $T_{xx}$ and $T_{yx}$ are around $0.55$. The measured magnitude is slightly lower than the simulated one, which is $0.565$. It is reasonable due to the measurement error and imperfect material properties of substrate.

\subsection{Comparisons}
Another chiral sample with the same configuration of the chiral circular polarizer except for the twisting angle $\alpha$ ($30^{\circ}$) was fabricated and measured for comparison.

\begin{figure}[!tbc]
\centering
\psfrag{c}[l][c][0.5]{$\bm{T_{xx}}$ (Sim.)}
\psfrag{e}[l][c][0.5]{$\bm{T_{xx}}$ (Meas. no gate)}
\psfrag{g}[l][c][0.5]{$\bm{T_{xx}}$ (Meas. gate)}
\subfigure[]{
\includegraphics[width=0.48\columnwidth]{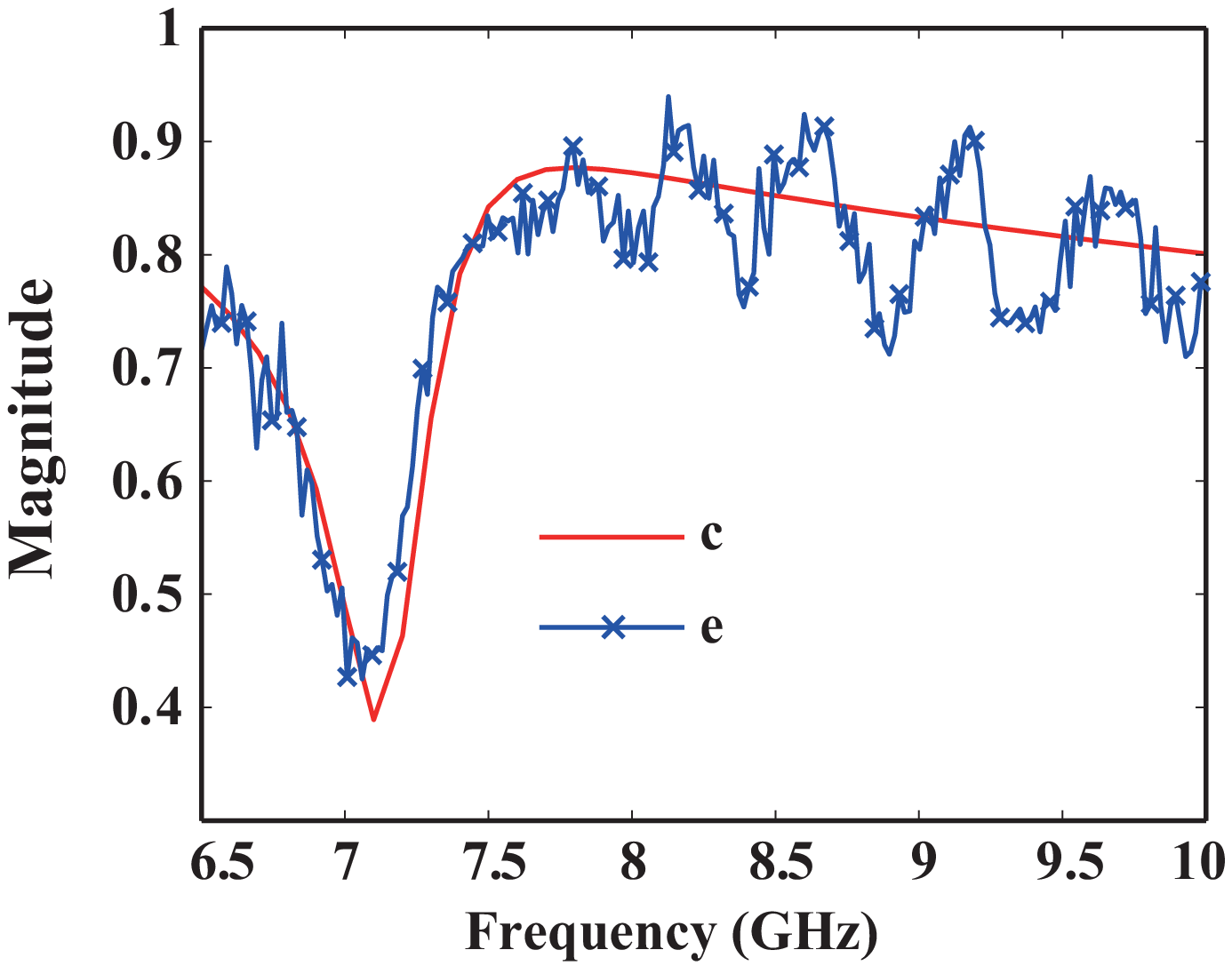}}
\subfigure[]{
\includegraphics[width=0.48\columnwidth]{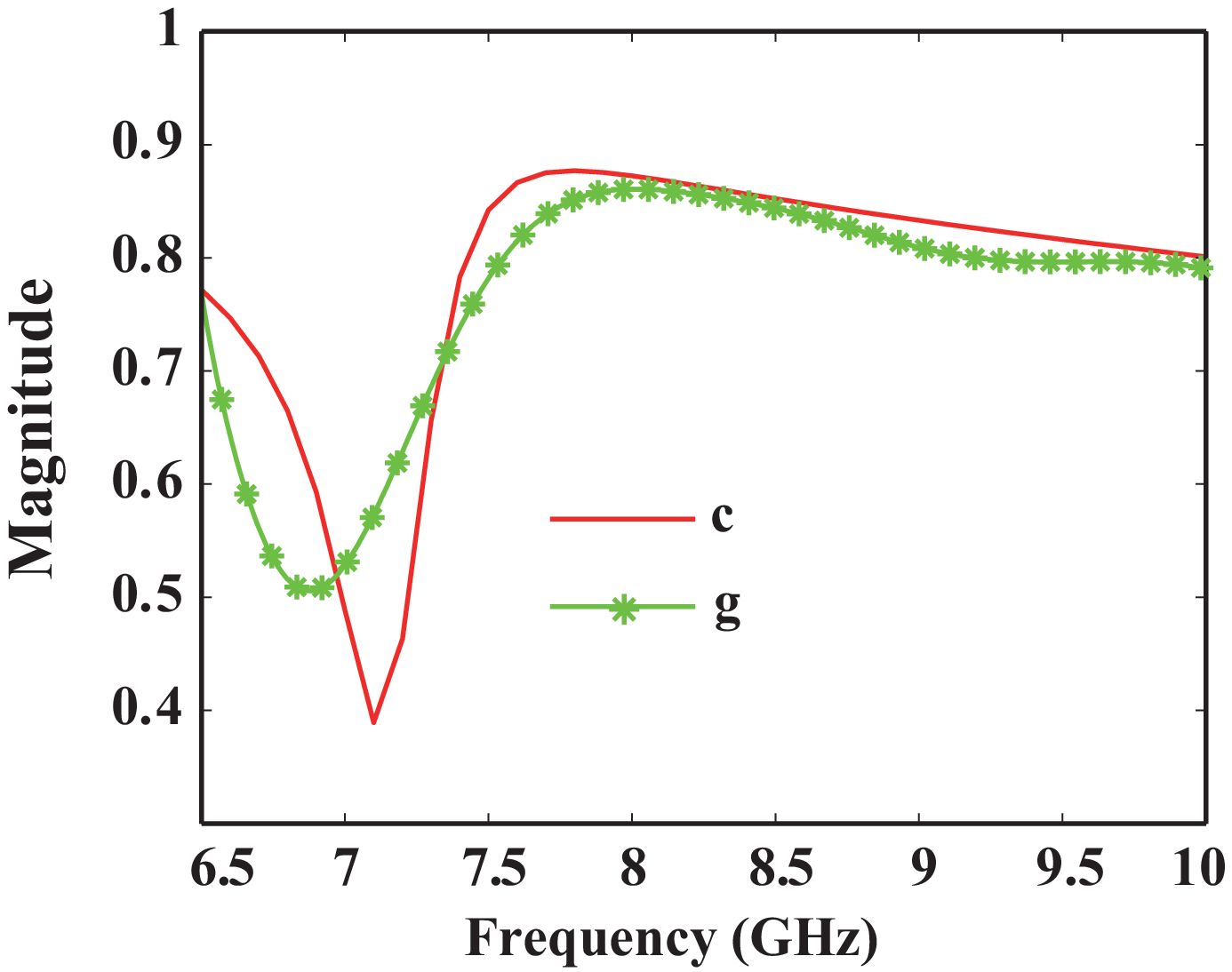}}
\caption{Simulation and experimental results of the modified chiral sample ($\alpha=30^\circ$). (a) simulation and measurement results without time-domain gate; (b) simulation and measurement results with time-domain gate.}
\label{gate}
\end{figure}

During the measurement, we found that the measured data was inaccurate when the time-domain gate was applied, as plotted in Fig.~\ref{gate}(b). It is known that sharp changes in frequency domain imply a broadband time-domain response. To recover the sharp response of the sample around $7.2$~GHz, time-domain information during a large time interval is needed. However, the desired time-domain gate avoiding the multiple reflections also filters out the useful information. Truncation of this time-domain response will smoothen and broaden the tip in frequency domain, which is consistent with the results in Fig.~\ref{gate}(b). Trend of the measurement result without using a time-domain gate follows that of the simulated one well but with ripples as shown in Fig.~\ref{gate}(a).

\begin{figure}[!tbc]
\centering
\psfrag{m}[c][c][0.5]{$\bm{\theta}$ \textbf{(deg)}}
\psfrag{n}[c][c][0.5]{$\bm{\eta}$ \textbf{(deg)}}
\psfrag{c}[l][c][0.5]{$\bm{\alpha}=90^\circ$ (Sim.)}
\psfrag{e}[l][c][0.5]{$\bm{\alpha}=30^\circ$ (Sim.)}
\psfrag{g}[l][c][0.5]{$\bm{\alpha}=90^\circ$ (Meas.)}
\psfrag{h}[l][c][0.5]{$\bm{\alpha}=30^\circ$ (Meas.)}
\psfrag{a}[c][c][0.5]{$\bm{\alpha}$ }
\subfigure[]{
\includegraphics[width=0.48\columnwidth]{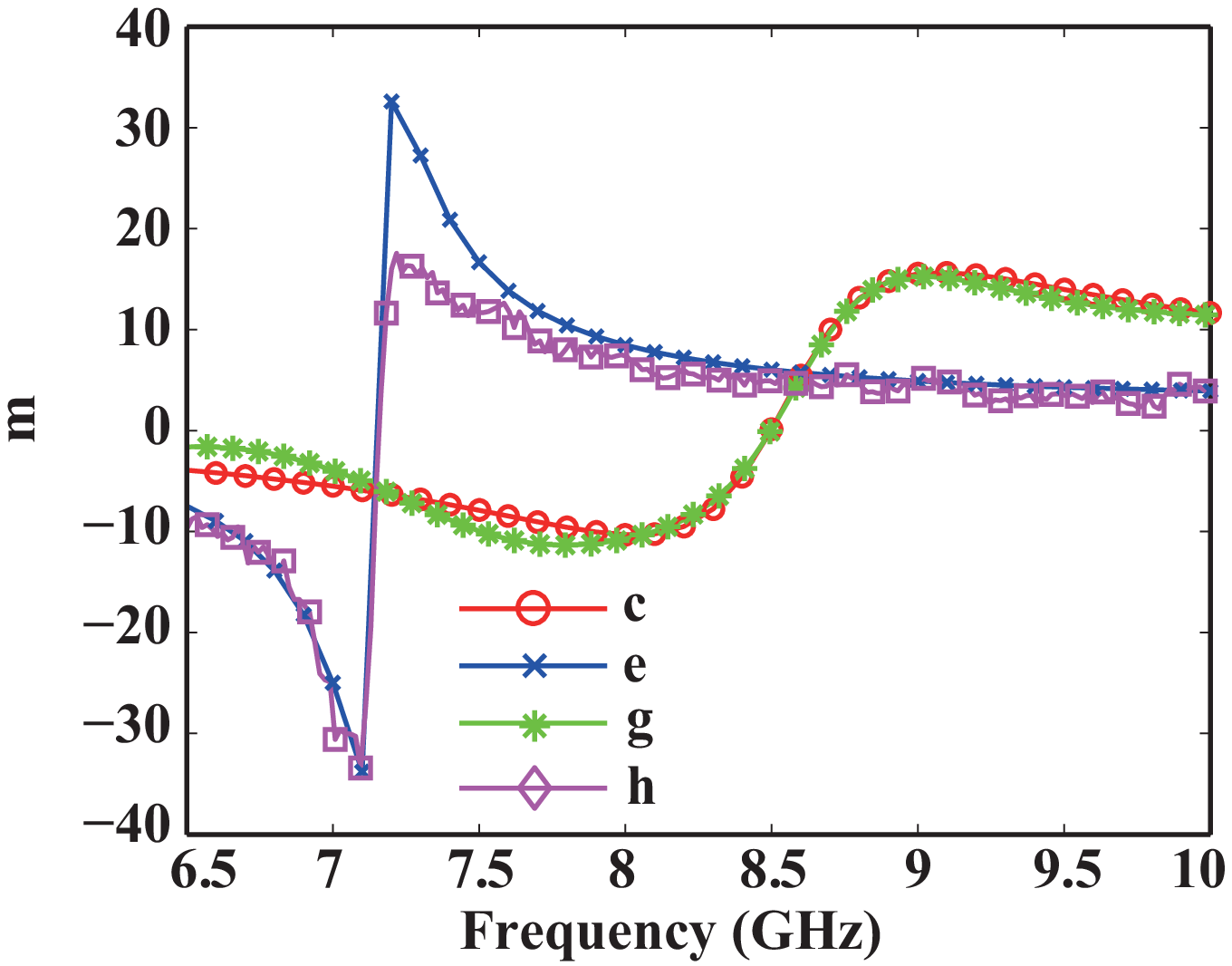}}
\subfigure[]{
\includegraphics[width=0.48\columnwidth]{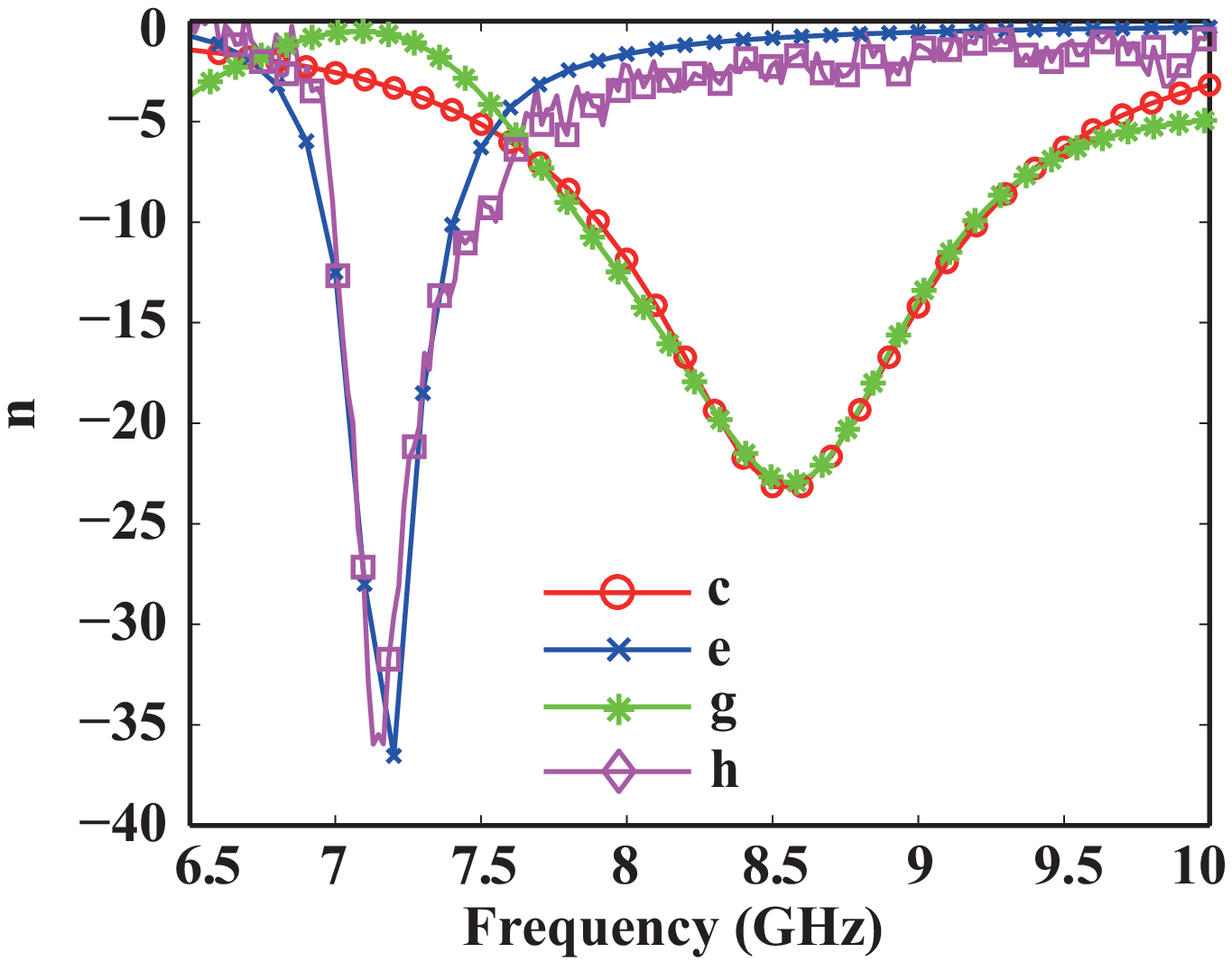}}
\caption{(a) Optical activity and (b) circular dichroism of the chiral circular polarizer ($\alpha=90^\circ$) and the modified chiral structure ($\alpha=30^\circ$).}
\label{60}
\end{figure}

Therefore, we abandoned the time-domain gate during the measurement of the chiral sample with $\alpha=30^\circ$. Polarization responses of the sample with $\alpha=30^\circ$ and the chiral circular polarizer are examined and compared both numerically and experimentally. Azimuthal rotation angle $\theta$ and ellipticity $\eta$ are plotted in Fig.~\ref{60}. When $\alpha=30^\circ$, both of the two parameters becomes larger. Good agreements can be observed between the simulation and experiment results. Effect of the signal multi-reflections between antennas and the board with $\alpha=30^\circ$ can be found in the graph.

\section{Supercell arrangement}

\begin{figure}[htbp]
\begin{center}
\subfigure[]{
\includegraphics[width=0.4\columnwidth]{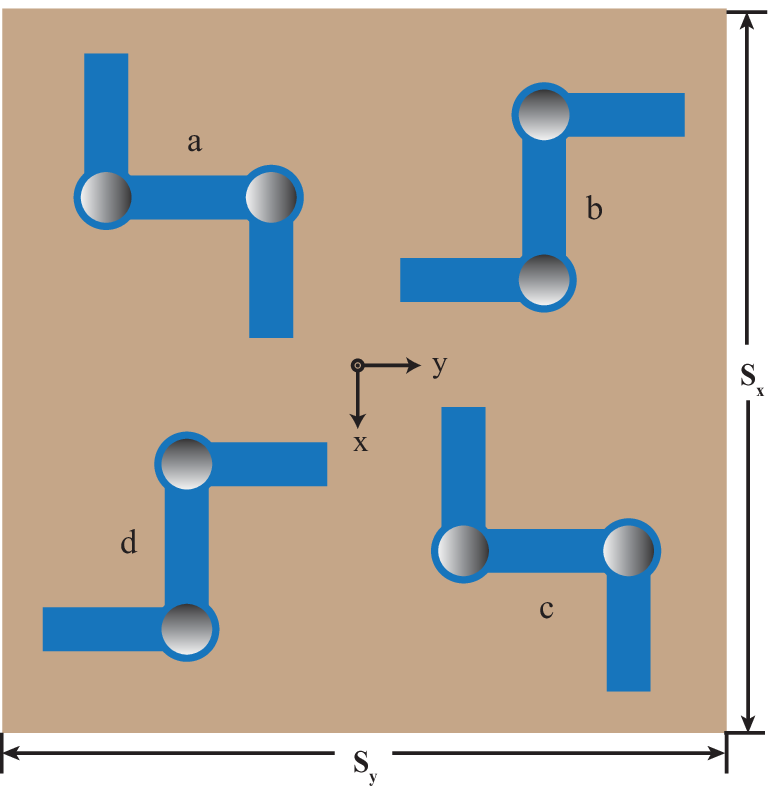}}
\subfigure[]{
\psfrag{m}[c][c][0.5]{$\bm{\theta}$ \textbf{(deg)}}
\psfrag{n}[c][c][0.5]{$\bm{\eta}$ \textbf{(deg)}}
\includegraphics[width=0.5\columnwidth]{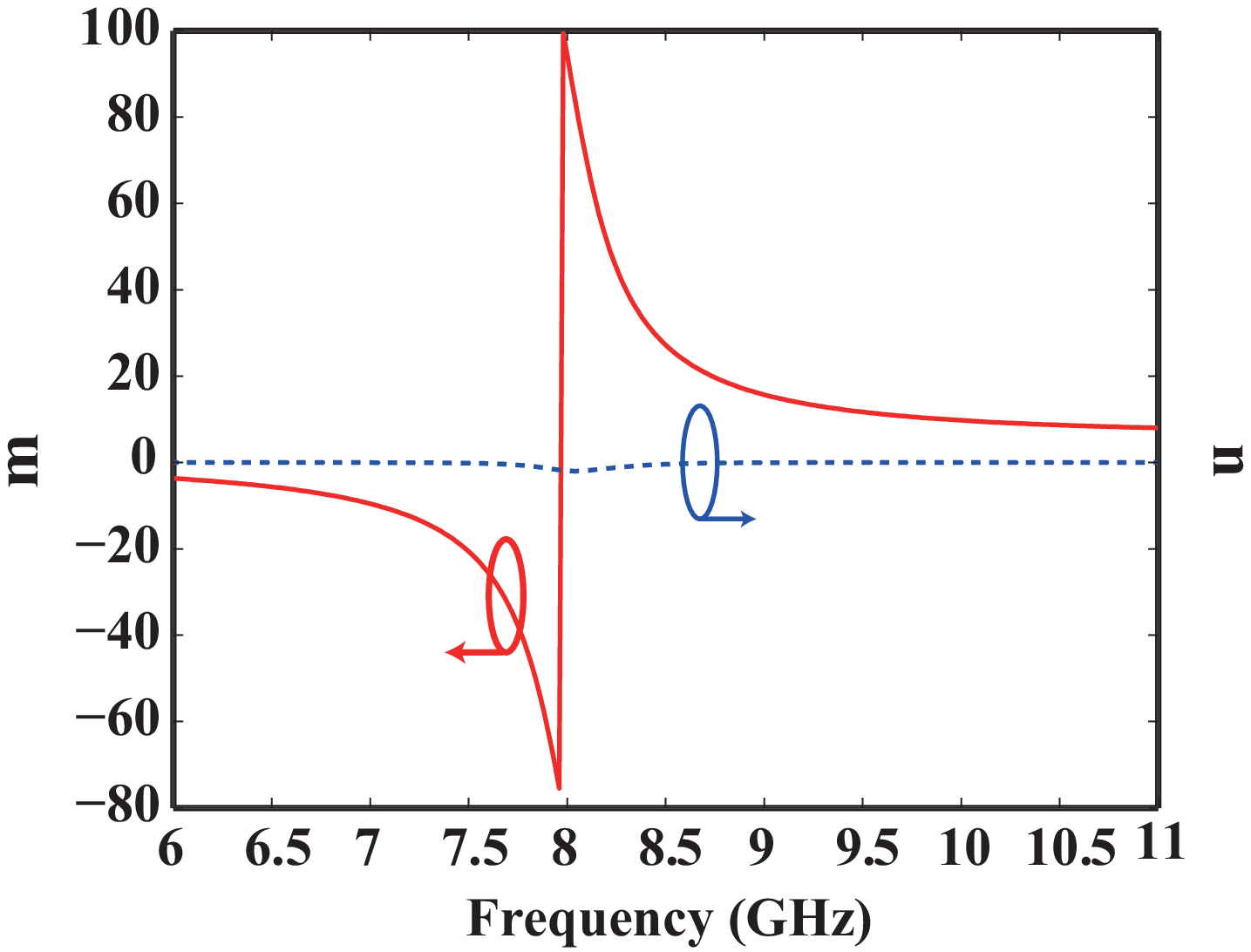}}
\caption{(a) schematic of the uniaxial supercell structure; (b) simulation results of the azimuthal rotation angle and ellipticity.}
\label{super}
\end{center}
\end{figure}

Till now, the chiral sample is sensitive to the polarization direction of normal incidence waves. We can achieve the isotropy under the normal incidence by arranging the omega-like particle in $C_4$ symmetry manner (See Fig.~\ref{super}). The four particles have the identical parameters as the chiral circular polarizer proposed in section IV. The supercell is periodic along the $x$ and $y$ directions with the periodicity of $13 \times 13$~cm$^2$. The supercell size is $0.4 \lambda_0 \times 0.4 \lambda_0$ at $9.2$~GHz. Therefore, it can be considered as a uniaxial structure for the normal incidence wave. Besides the common features of the $C_4$ symmetric particle, i.e. $T_{yx}=-T_{xy}$ and $T_{yy}=T_{xx}$, another feature can be found by the simulation results in Fig.~\ref{super}(b). The ellipticity is very low with a maximum value of $2^{\circ}$ around $8$ GHz while the azimuthal rotation angle is $90^{\circ}$. Nearly pure cross-polarized wave is generated at $8$ GHz.

\section{Conclusions}

In summary, to explore a strong polarization control capability, we proposed and systematically studied a 3D omega-like chiral structure. The transmitted polarization states from the chiral structure are highly tunable, which is characterized by a large range of azimuthal rotation angle and ellipticity. Based on the proposed chiral particle, we also successfully realized chiral circular polarizer, through which the linear polarized wave can be converted to the RCP or LCP waves. Experimental results show good agreements with the simulated ones. Finally, we developed a uniaxial chiral slab.

\section*{Acknowledgment}

This work was supported in part by the Research Grants Council of Hong Kong (GRF 716713, GRF 17207114, and GRF 17210815), NSFC 61271158, Hong Kong ITP/045/14LP, Hong Kong UGC AoE/P¨C04/08, the Collaborative Research Fund (Grants C7045-14E) from the Research Grants Council of Hong Kong Special Administrative Region, China, and Grant CAS14601 from CAS-Croucher Funding Scheme for Joint Laboratories.

\bibliographystyle{IEEEtran}
\bibliography{reference}

\ifCLASSOPTIONcaptionsoff
  \newpage
\fi

\end{document}